\documentclass[prb,twocolumn,groupeaddress]{revtex4-2}

\usepackage{amsmath}
\usepackage{graphicx}
\usepackage[colorlinks=true, allcolors=blue]{hyperref}

\begin{document}

\title{Electronic instability, layer selectivity and Fermi arcs in La$_3$Ni$_2$O$_7$}
\author{Frank Lechermann}
\affiliation{Institut f\"ur Theoretische Physik III, Ruhr-Universit\"at Bochum,
  D-44780 Bochum, Germany}
  \author{Steffen B\"otzel}
  \affiliation{Institut f\"ur Theoretische Physik III, Ruhr-Universit\"at Bochum,
  D-44780 Bochum, Germany}
\author{Ilya M. Eremin}
\affiliation{Institut f\"ur Theoretische Physik III, Ruhr-Universit\"at Bochum,
  D-44780 Bochum, Germany}

\pacs{}
\begin{abstract}
Using advanced dynamical mean-field theory on a realistic level we study the normal-state correlated electronic structure of the high-pressure superconductor La$_3$Ni$_2$O$_7$ and compare the features of the conventional bilayer (2222) Ruddelsden-Popper crystal structure with
those of a newly-identified monolayer-trilayer (1313) alternation. Both structural cases display
Ni-$d_{z^2}$ flat-band character at low-energy, which drives an electronic instability with a wave vector ${\bf q_{\rm I}}=(0.25,0.25,q_z)$ at ambient pressure, in line with recent experimental findings. The 1313 electronic structure exhibits significant layer selectivity, rendering especially the monolayer part to be Mott-critical. At high pressure, this layer selectivity weakens and the 1313 fermiology displays arcs reminiscent to those of high-$T_c$ cuprates. In contrast to dominant inter-site self-energy effects in the latter systems, here the Fermi arcs are the result of the multiorbital and multilayer interplay within a correlated flat-band scenario.
\end{abstract}

\maketitle

\section{Introduction} 
Last year's discovery of high-temperature superconductivity in La$_3$Ni$_2$O$_7$ (La-327)
at high pressure~\cite{sun23} added a fascinating new chapter to the still juvenescent field of superconducting nickelates~\cite{li19,li20,zeng20,pan21}. It originated several follow-up
studies, both from experiment~\cite{zhangsu23,houyang23,wangwangpoly23,zhouguo23,wangwang23,liguo23}
as well as from theory~\cite{luohu23,zhanglin23,Zhang2023second,lechermann23,shilenko23,chen-jul23,sakakibara24,christiansson23,yangwangwang23,lupan23,liumei23,yangzhang23,xing-zhou24,jianghuo23,luobiao23,oh24,boetzel24,ryee2024quenched,Savrasov2024}. Despite efforts to directly tackle the superconducting
phase, understanding the nature of the superconductivity in La-327 is likely related to the quest of 
decoding the correlated normal-state electronic structure. 

In a standard setting, La-327 crystallizes in the $n=2$ bilayer structure (2222) of the
Ruddelsden-Popper series La$_{n+1}$Ni$_n$O$_{3n+1}$~\cite{lin99} (see Fig.~\ref{fig1}a),
with a nominal Ni$^{2.5}$ oxidation state~\cite{drennan82,gervais85,sav88,zhang94}.
Ligand-hole physics was however predicted~\cite{lechermann23,shilenko23,chen-jul23} to yield a Ni$(3d^8)$ rather than the formal Ni$(3d^{7.5})$ filling, as recently supported by electron energy loss spectroscopy~\cite{dong-dec23} and resonant inelastic X-ray scattering (RIXS)~\cite{chenchoi-jan24}. Hence two holes/electrons reside in the twofold Ni-$e_g$
orbital sector. At ambient pressure, metallic response with resistance
anomalies at $T_{\rm a1}\sim 140-150$\,K~\cite{liusun22} and
$T_{\rm a2}\sim 110$\,K~\cite{zhang94,sreedhar94,kobayashi96,liusun22} is measured. Recent experimental
studies connect $T_{\rm a2}$ to a partial Fermi-surface removal~\cite{liuhuo-jul23}, and
$T_{\rm a1}$ to a spin density-wave (SDW) transition~\cite{chenliu-nov23,chenchoi-jan24,danzhuo-feb24,khasanov24} with an in-plane component ${\bf q_{\rm DW}}=(0.25,0.25)$ of the ordering wave
vector, and presumably antiferromagnetic ordering between the NiO$_2$ layers, 
as determined by RIXS~\cite{chenchoi-jan24}. 

On top of these intriguing behavior with temperature, a competing structural motif
of alternating monolayer-trilayer kind (1313) has recently been identified in
La-327~\cite{chenzhang24,puphal23,wangchenruther24} (see Fig.~\ref{fig2}a). At the
present stage, it seems that both the 2222 and the 1313 motif are realized in the compound, with area
sizes depending on the crystal-growth conditions. Which electronic-state behavior, including
superconductivity at high pressure, is linked to either of these two allotropes, remains elusive
so far. Low-$T$ angle-resolved photoemission spectroscopy (ARPES)
measurements~\cite{yangsun23,abadi24} {\sl well below} $T_{\rm a1}$ find a flat band beneath
the Fermi level for both structures, albeit the flat-band location is somewhat
deeper in the 2222 ($\sim 50$\,meV) than in the 1313 ($\sim$ 25 meV) case. Notably, this flat band
is still weakly crossing the Fermi level from conventional density functional theory (DFT)
calculations. Usually, DFT is not too bad for the principle interacting Fermi-surface topology and importantly, that result still describes the low-energy dispersions in a high-symmetry phase,
i.e. formally {\sl above} the SDW transition at $T_{\rm a1}$. On the other hand, DFT+U~\cite{yangsun23}
and related~\cite{wangwangzhang24} approaches shift this flat band indeed below the
Fermi level already in the high-symmetry phase. On a purely technical level, this is not too
surprising: DFT+U by design always ``tries'' to open gaps through completely filling/emptying weakly
doped/filled bands.

In this work, the 2222 and 1313 structural motifs are compared and several key features of the La-327 normal state are uncovered from a first-principles many-body perspective. The $T_{\rm a1}$ anomaly originates from a SDW transition, mainly driven by highly-renormalized Ni-$d_{z^2}$ derived electronic states. The aforementioned placement of the flat band beneath the Fermi level at low $T$ is therefore a result of the ordering transition.
A strongly layer-selective character in the 1313 structure gives furthermore rise to significant
differences in the correlated-electron behavior between the mono- and trilayer segments. Last but
not least, a characteristic Fermi-arc feature is found in the high-pressure 1313 phase, resulting 
from the strongly-correlated interplay of the multiorbital/layer electronic structure, with the concomitant, now low-energy stuck, flat-band feature.

\section{Theoretical Approach}
We employ the charge self-consistent~\cite{grieger12} DFT+sicDMFT framework~\cite{lechermann19}, building up on the original DFT+DMFT method (see e.g.~\cite{kotliar06} for a review). The Ni sites are quantum impurities in dynamical mean-field theory (DMFT) and Coulomb interactions on oxygen enter via the self interaction correction (SIC) on a
pseudopotential level~\cite{korner10}. The DFT part uses a mixed-basis pseudopotential scheme~\cite{elsaesser90,lechermann02,mbpp_code} in
the local-density approximation (LDA). The SIC is applied to the O$(2s,2p)$ orbitals via weight factors $w_p$. While the $2s$ orbital is fully corrected with $w_p=1.0$, the choice~\cite{korner10,lechermann19,lechermann23,chen22} $w_p=0.8$
is used for O$(2p)$. Continuous-time quantum Monte Carlo in hybridization
expansion~\cite{werner06} as implemented in the TRIQS code~\cite{parcollet15,seth16} solves the DMFT problem. A five-orbital Slater-Hamiltonian, parameterized by Hubbard $U=10$\,eV and
Hund exchange $J_{\rm H}=1$\,eV~\cite{lechermann20-1}, governs the correlated subspace defined by Ni$(3d)$ projected-local orbitals~\cite{amadon08,aic09}. All calculations aim for a paramagnetic regime.

Further more technical details are as follows. Concerning DFT, a $5$$\times$$5\times$$5$ for $2222-Amam$, a $3$$\times3$$\times$$3$ for $1313-Fmmm$ and a $7$$\times$$7\times$$2$ $k$-point mesh for $1313-P4/mmm$ (see below) is utilized. The plane-wave cutoff energy is generally set to $E_{\rm cut}=15$\,Ry. Local basis orbitals are introduced for La$(5d)$, Ni$(3d)$ and O$(2s,2p)$. The role of possible spin-orbit effects is neglected in the crystal calculations, but the pseudopotentials are generated with including spin-orbit coupling.
The DMFT correlated subspace on each Ni site is governed by a rotational-invariant five-orbital Slater Hamiltonian, applied to the Ni$(3d)$ projected-local orbitals. The projection is performed on the Kohn-Sham (KS) bands above the respective bands of dominant O$(2s)$ character for each structural case. The projection window spans (68,136,68) bands for ($2222-Amam,1313-Fmmm,1313-P4/mmm$), including the KS states of dominant Ni$(3d)$ and O$(2p)$, as well as one additional KS band for each La site in the primitive cell. For the solution of the multi-site impurity problems, up to $5\cdot 10^9$ Monte-Carlo sweeps are performed to reach convergence. A Matsubara mesh of 1025 frequencies is used to account for the given temperature regime $T\ge 100$\,K. For the analytical continuation from Matsubara space onto the real-frequency axis, the Maximum-entropy method~\cite{jar96} is used for the {\bf k}-integrated spectra (by continuation of the Bloch Green’s function) and the Pad{\'e} method~\cite{vid77} is employed for the {\bf k}-resolved spectra (by continuation of the local self-energies). The fully-localized-limit double-counting correction scheme~\cite{anisimov93} is applied.

In the result section, different forms of spectral functions are discussed. The projected spectral function $A_{\rm proj}$ is derived from the interacting Bloch Greens function $G_{\rm bl}$ of the full DFT+sicDMFT Hilbert space~\cite{grieger12}. Thus this projected spectral function covers the full hybridization effects in the interacting lattice regime. The total spectral function $A_{\rm tot}$ represents the sum of the site- and orbital-resolved $A_{\rm proj}$. In that sense, $A_{\rm tot}$ and $A_{\rm proj}$ are the interacting counterparts of the total as well as site-and orbital-resolved density of states (DOS) of standard KS-DFT. Note also that aside from the purely interacting viewpoint, the effect of temperature in DFT+(sic)DMFT is not only to change the occupational features of the spectrum according to Fermi-Dirac statistics as in finite-$T$ DFT. Here, the temperature introduces also a coherence scale for the electronic excitations, above which those cease to exist.

The bare susceptibility is computed as $\chi_0({\bf q})\sim {\rm Tr}\,G({\bf k})G({\bf k}+{\bf q})$, whereby
$G$ is the interacting DFT+sicDMFT (Bloch) Green's function. The trace ${\rm Tr}$ is over band indices and Matsubara frequencies. Individual calculations are performed on a two dimensional 20$\times$20 mesh in reciprocal space for different $k_z$ values, respectively, and accordingly summed. Correspondingly, finite $q_z$ values can be attained by connecting meshes with different $k_z$ values.

\section{Results}
\begin{figure}[t]
      \includegraphics[width=\linewidth]{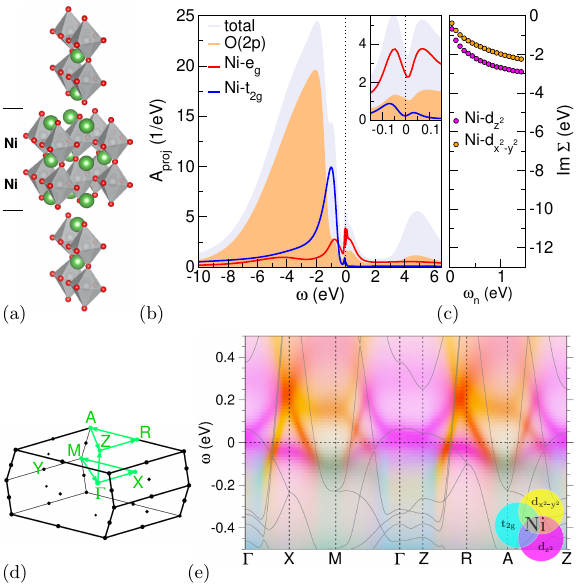}
      \caption{Correlated electronic structure of 2222 La-327 at ambient pressure ($T=150$\,K).
        (a) Illustration of the $Amam$ crystal structure: La (green), Ni (grey) and
        O (red). (b) Total and projected spectral function (inset: low-energy blow up).
        (c) Imaginary part of the Ni-$e_g$ self-energies as a function of Matsubara frequencies $\omega_n$.
        (d) Brillouin zone (BZ) with high-symmetry lines. (e) Low-energy {\bf k}-resolved spectral
        function in fatspec representation, i.e. color differentiation marks the varying
        Ni$(3d)$ orbital weights. Grey lines mark the bare DFT electronic dispersion.}\label{fig1}
\end{figure}	
Let us start with the correlated electronic structure of the 2222 La-327 structure at ambient pressure
and $T=150$\,K, i.e. close to the experimental $T_{\rm a1}$, shown in Fig.~\ref{fig1}. The crystal
data of the orthorhombic $Amam$ structure (two-formula-unit primitive cell) is taken from
Ref.~\onlinecite{lin99}. Note that all Ni sites are equivalent by symmetry. The {\bf k}-integrated
spectrum displayed in Fig.~\ref{fig1}b highlights once again~\cite{lechermann23} the Ni-$e_g$
dominance at lower energy. A pseudogap signature of $~\sim 100$\, meV width holds close to the
Fermi level $\varepsilon_{\rm F}$.
\begin{figure}[t]
  \includegraphics[width=\linewidth]{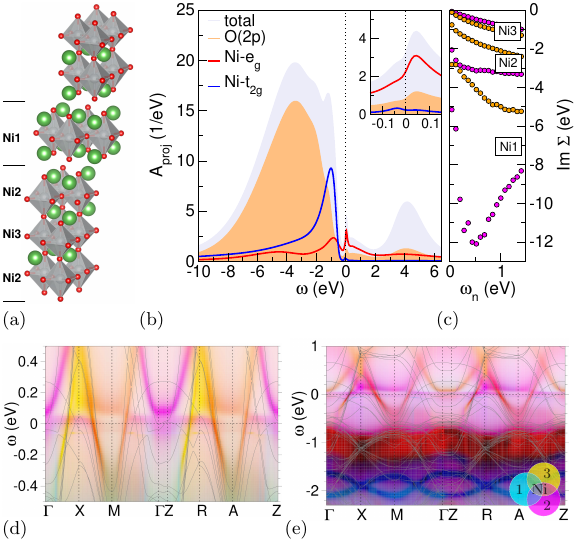}
  \caption{Correlated electronic structure of 1313 La-327 at ambient pressure ($T=150$\,K).
          (a) Illustration of the $Fmmm$ crystal structure.
    (b) Total and projected spectral function (inset: low-energy blow up), normalized
     to a two-formula-unit primitive cell.
        (c) ${\rm Im}\,\Sigma(i\omega_n)$ of Ni1,2,3-$e_g$,
        respectively.
        (d) Low-energy spectral function $A({\bf k},\omega)$ in
        fatspec representation (color code as in Fig.~\ref{fig1}e).
        (e) $A({\bf k},\omega)$ in larger energy window
        and fatspec now according to Ni1,2,3.  Grey lines in (d,e) mark the bare DFT electronic dispersion.}
        \label{fig2}
\end{figure}	
\begin{table}[b]
\begin{ruledtabular}
  \begin{tabular}{l|cccc}
              &  2222   &            & 1313     &    \\
              &    Ni   & Ni1        &  Ni2      & Ni3 \\[0.05cm] \hline
$d_{z^2}$     & 1.07    & 1.12,  1.07 & 1.05, 1.07 & 0.89, 0.92    \\
              &         & (1.27, 1.50) & (1.20, 1.17) & (1.06, 1.04) \\[0.1cm]
$d_{x^2-y^2}$ & 1.02    & 1.13,  1.06 & 0.98, 1.02 & 0.99, 1.04    \\
              &         & (1.20, 1.18) & (1.19, 1.02) & (1.12, 1.07) \\[0.1cm]
$3d$ total    & 8.06    & 8.23,  8.11 & 8.01, 8.05 & 7.85, 7.92     \\
              &         & (8.30, 8.58) & (8.24, 8.03) & (8.08, 8.00) \\
\end{tabular}
\end{ruledtabular}
\caption{Orbital- and layer-resolved Ni$(3d)$ fillings in the 2222 and 1313 structure. Both values for Ni1,2,3 in each row refer to ambient and high pressure phases, respectively. The values in braces refer to the respective LDA filling.}
\label{tab:fill}
\end{table}
As expected from the ligand-hole physics, Tab.~\ref{tab:fill} renders the Ni$(3d^8)$ fillings with one electron in each Ni-$e_g$ orbital. An average O$(2p)$ occupation of
$n_{2p}=5.62$ is obtained, i.e. there are about 0.4 holes per oxygen. The Ni-$d_{z^2}$ orbital
shows the more prominent self-energy behavior in Fig.~\ref{fig1}c, yet a clear Fermi-liquid (FL)
regime with a close-to-linear ${\rm Im}\,\Sigma(i\omega_n)$ behavior for small $\omega_n$~\cite{Tremblay2012} cannot
be identified for either of both Ni-$e_g$ orbitals. On a more formal level, a low-Matsubara-frequency
fit to $B_0\,\omega_n^{\alpha}$ leads to exponents much lower than the FL value $\alpha=1$.
A formal brute-force FL fit results in effective masses $m^*\gtrsim 8$. The non-Fermi-liquid
behavior is most likely associated with the Ni-$d_{z^2}$ flat-band physics originating around the $\Gamma$-point of the BZ (cf. Fig.~\ref{fig1}d), as visualized in Fig.~\ref{fig1}e. Note the strong shift of $d_{z^2}$ spectral weight towards
$\varepsilon_{\rm F}$, e.g. at X and from above at $\Gamma$, even not connected to the original
flat DFT dispersion at $\Gamma$. We will come back to this important result when we discuss the
electronic instability inherent to this correlated state.

Next we discuss the corresponding 1313 correlated electronic structure at ambient pressure
(see Fig.~\ref{fig2}). Here, we utilize the crystal data by Puphal {\sl et al.}~\cite{puphal23}, describing
orthorhombic $Fmmm$ symmetry. The primitive cell consists
of a four-formula-unit cell, with three inequivalent Ni sites Ni1 (2 ions), Ni2 (4 ions) and
Ni3 (2 ions). From Fig.~\ref{fig2}a, Ni1 is associated with the monolayer and Ni2(3) with
the outer(inner) layer of the trilayer segment. Not surprisingly, the total and projected $A(\omega)$ in
Fig.~\ref{fig2}b look rather similar to the 2222 case, with minor differences
in view of the location of the O$(2p)$ main peak and the decoupling from Ni-$t_{2g}$. At the same time,
the low-energy pseudogap has weakened to a mere shoulder-like feature in 1313. Furthermore, stronger differentiation
appears on the local level, where a substantial Ni-$e_g$ layer selectivity is observed in the
self-energy (Fig.~\ref{fig2}c) and the Ni$(3d)$ occupation (Tab.~\ref{tab:fill}). The Ni1 position from the
monolayer is effectively Mott-critical, seemingly in line with the Mott-insulating state of bulk La$_2$NiO$_4$~\cite{lechermann22-2}. The outer Ni2 from the trilayer is
somewhat less correlated, but again not ``in good FL shape''. In contrast, the inner-layer Ni3
self-energies (note the Ni-$e_g$ internal hierarchy change) are even smaller than those for Ni in
the 2222 structure. While still not perfectly FL-like, values of $\alpha\sim 0.8$ point to
near-quasiparticle behavior with formal $m^*\sim 2(3)$ for $d_{z^2}(d_{x^2-y^2})$. This transfer
of correlation strength from Ni3 site to (Ni2,Ni1) sites is accompanied by electron transfer in the same
direction: Ni3-$d_{z^2}$ is effectively about 10\% hole doped. On the other hand, the ligand-hole
amount remains unchanged with $n_{2p}=5.62$. Especially the substantial hole doping on the Ni3 site and the Ni1-$d_{z^2}$ filling are qualitatively not well captured from a sole LDA perspective for the fillings (see Tab.~\ref{tab:fill}). Strong correlation effects have thus to be treated on a reasonable level to account for the actual charge transfers.

The {\bf k}-dependent 1313 electronic spectrum in Fig.~\ref{fig2}d is again similar to the 2222 case, but the
Ni-$d_{z^2}$ states shift towards $\varepsilon_{\rm F}$ is reduced, especially close to the X-point of the BZ from below. In
fact, the $\Gamma$-X $d_{z^2}$ weight close to the Fermi level has even a slightly reversed up-below
appearance. Figure~\ref{fig2}e additionally shows the states in reciprocal space according to a layer-resolved fatspec representation,
marking the expected shift of Ni1 weight to higher energies. The low-energy regime is still shared
by Ni2 and Ni3, a very obvious sole Ni3 dispersing part cannot readily be identified.
\begin{figure}[t]
      \includegraphics[width=\linewidth]{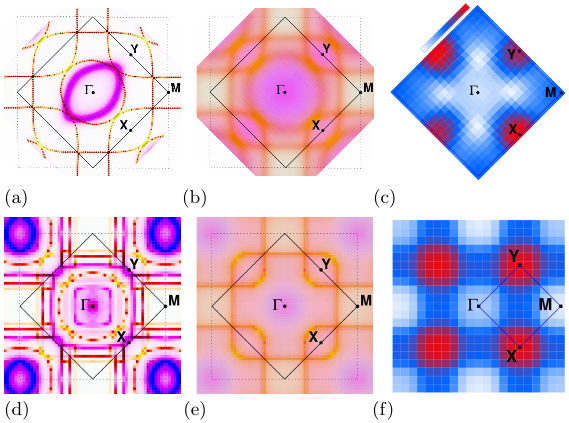}
      \caption{Fermiology for $k_z=0$ at ambient pressure and $T=150$\,K
        for 2222 (a-c) and 1313 (d-f).
        (a,d) DFT Fermi surface and (b,e) DFT+sicDMFT Fermi surface, both in fatspec representation
        with color coding as in Fig.~\ref{fig1}e. Full black lines: matching BZ, dashed black lines:
        pseudo-tetragonal BZ. (c,f) static bubble susceptibility $\chi_0({\bf q})$.}\label{fig3}
\end{figure}	

In order to connect the spectral findings to the experimental SDW instability at $T_{a1}$, we proceed with an assessment of the fermiology and the linear response function. The
$k_z=0$ FSs in Fig.~\ref{fig3} are plotted together with the tailored BZ (full lines) and the
associated pseudo-tetragonal BZ (dashed lines) of a single-Ni in-plane unit cell. Therefore, e.g.
the X point has (0.25,0.25) absolute-value units in the pseudo-tetragonal setting.
The 2222 DFT-FS shown in Fig.~\ref{fig3}a has three dominant sheets: two strongly-hybridized
Ni-$e_g$ sheets, one circular electron-like around $\Gamma$ and one hole-like in cuprate manner,
as well as a peculiar orthorhombically-distorted Ni-$d_{z^2}$ sheet around $\Gamma$ stemming from
the notorious flat band. For completeness, the corresponding 1313 DFT-FS is shown in Fig.~\ref{fig3}d,
and while the averaged features are similar, things are blurred out by larger-cell multiplicities.
For instance, notably an additional tiny $\Gamma$ point pocket from Ni-$d_{z^2}$ is observable. Upon adding correlations, common tendencies are clearly observed. In particular, one finds that the Ni-$d_{z^2}$ derived flat band parts become very
incoherent and appear smeared. The only dispersing parts are of hybridized Ni-$e_g$ (bonding) character, present for both 1313 and 2222 structures (see Fig.~\ref{fig3}(b,e)). Note that the inner electron-like sheet of that kind becomes very weak in the 1313 case. The sizes of the respective sheets also change, but the Luttinger theorem is of limited relevance in a non-Fermi-liquid regime.

Computing the bare bubble susceptibility $\chi_0({\bf q})\sim {\rm Tr}\,G({\bf k})G({\bf k}+{\bf q})$
from the interacting DFT+sicDMFT Green's function $G$ leads to a common picture for a principle
${\bf q}$-dependent instability in both structural types. The static $\chi_0$ has maxima at ${\bf q_{\rm I}}$ located at the X,Y points, consistent with the experimentally deduced in-plane ordering wave vector ${\bf q_{\rm SDW}}=(0.25,0.25)$. Interestingly, the orbital content at ${\bf q_{\rm I}}$ is dominated by Ni-$d_{z^2}$, similar to what was deduced at ${\bf q_{\rm SDW}}$ from RIXS~\cite{chenchoi-jan24}. Thus while within a weak-coupling picture, the origin of the weakly incommensurate peak near  ${\bf q_{\rm SDW}}$ is connected to the bonding-antibonding scatterings of mostly Ni-$d_{x^2-y^2}$ derived bands~\cite{wangwangzhang24}, in our strong-coupling picture the flat regions of Ni-$d_{z^2}$ character around $\Gamma$ and X of the BZ (cf. Fig.~\ref{fig1}e), give rise to the commensurate peaks at ${\bf q_{\rm I}}={\bf q_{\rm SDW}}=(0.25,0.25)$.
\begin{figure}[t]
      \includegraphics[width=\linewidth]{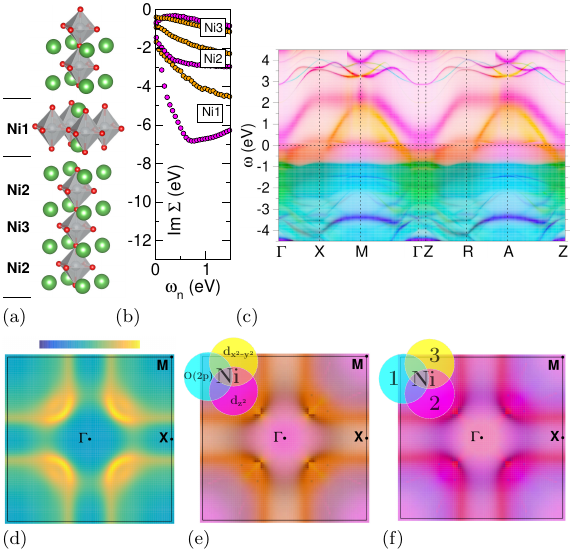}
      \caption{Correlated electronic structure of 1313 La-327 at high pressure ($T=100$\,K)
          (a) Illustration of the $P4/mmm$ crystal structure.
          (b) ${\rm Im}\,\Sigma(i\omega_n)$ of Ni1,2,3-$e_g$, respectively.
        (c) Spectral function $A({\bf k},\omega)$ in
        fatspec representation (color code identical to Fig.~\ref{fig1}e).
        (d-f) $k_z=0$ fermiology: FS by (d) intensity, (e) fatspec according to orbital content
        and (f) fatspec according to Ni layer content.}\label{fig4}
\end{figure}	
 It follows, that the corresponding flat-band feature observed in ARPES {\sl beneath} $\varepsilon_{\rm F}$ is a result of this instability, once the partial
gap opening has taken place. In other words, recent ARPES measured the ordered state below $T_{a1}$.
Note that quantitatively, $\chi_0({\bf q_{\rm I}})$ turns out smaller for 1313 than for 2222, which might explain the reduced split-off from the Fermi level in the former case as seen by ARPES. Moreover when allowing for finite $q_z$ in the computation of $\chi_0({\bf q})$, the in-plane location of ${\bf q_{\rm I}}$ remains robust, but the values are slightly enhanced. This is in line with previous theoretical findings of an enhancement of the odd-part susceptibility in La-327~\cite{boetzel24}. Yet the complete nature of the ordered state remains still open from the present investigation. Possible ordering patterns are discussed in Refs.~\onlinecite{chenchoi-jan24,khasanov24}. Vertex corrections to the susceptibility beyond the bubble~\cite{boehnke12,boehnke18,strand19} are tough to include for the given demanding multi-site/orbital problem. Yet those usually strengthen the effect of the
local-orbital correlations, and since those are here most
substantial for Ni-$d_{z^2}$, we assume that inclusion of vertex
corrections mainly enhances the strength of the Ni-$d_{z^2}$ driven
instability.
At high pressure, we expect the ${\bf q_{\rm SDW}}$ instability to be supressed, keeping the flat-band at the Fermi level~\cite{lechermann23,lupan23} to finally drive superconductivity. 

To inaugurate this high-pressure flat-band physics also in the context of the 1313 structural motif, Fig.~\ref{fig4} displays the correlated electronic structure at 16 GPa. The crystal data stems again from Ref.~\onlinecite{puphal23}, describing now tetragonal $P4/mmm$ symmetry,
i.e. NiO$_6$ octahedral tilts are absent (see Fig.~\ref{fig4}a). Figure~\ref{fig4}b shows that the
strength of the Ni1,2,3-$e_g$ self-energy differentiation is reduced with pressure, in line with
a reduction of the inter-layer charge transfers (see Tab.~\ref{tab:fill}). Also the Mottness of the
monolayer segment (Ni3) is weakened compared to the ambient system. However, the non-FL character
within the trilayer segment (Ni2,Ni3) is strongly enhanced, as readily observable from the
pathological behavior of ${\rm Im}\,\Sigma(i\omega_n)$ for small Matsubara frequencies. It originates
from strong quantum fluctuations initiated by the Ni-$d_{z^2}$ flat band stuck to the Fermi level
(cf. Fig.~\ref{fig4}c). Overall, the spectral weight close to $\varepsilon_{\rm F}$ is of hybridized
Ni-$e_g$ character, with nominally higher $d_{x^2-y^2}$ intensity.
Surprisingly, the interacting Fermi surface shown in Fig.~\ref{fig4}d displays Fermi-arc structure,
originally known from the pseudogap phase of high-$T_{\rm c}$ cuprates (see e.g. Refs.~\onlinecite{damascelli03,keimer15} for reviews). This means, the spectral intensity
is much higher for ``arcs'' perpendicular to the nodal $\Gamma$-M direction, whereas the intensity
appears blurred out in the anti-nodal direction towards $\Gamma$-X.
In cuprates, due to the effective one-band scenario and from a strong-coupling perspective, Fermi arcs are believed result from explicit {\bf k}-dependence of the electronic self-energy $\Sigma$~\cite{parcollet04,lichtkat00}. Yet such an explicit dependence is here neglected in our single-site DFT-sicDMFT approach.
By inspecting the orbital- and layer-resolved nature of the Fermi surface in Fig.~\ref{fig4}e,f, one may conclude that the robust spectral intensity within the arcs is based on a coherent effort over all
orbital/layer sectors. On the other hand, once this joint effort is lost away from the arcs, the correlated flat band is effective in dissolving spectral coherence. Based on the color coding (see inset), the dominant part of the arcs appears brown/dark in Fig.~\ref{fig4}e, which matches with sizable contributions from all examined orbital sectors (O$(2p)$: cyan, Ni-$d_{x^2-y^2}$: yellow, Ni-$d_{z^2}$: pink) from overlaying colors. The stronger lightbrown coloring, following more or less the original Fermi-surface sheet, amounts to missing strong contributions from O$(2p)$ and marks dominant Ni-$d_{x^2-y^2}$ character. In Fig.~\ref{fig4}f,
similar dark coloring for the key part of the arcs is observable,
now connecting to relevant contributions from all layers. And the
dominant pink coloring following again the original Ni-$d_{x^2-y^2}$ dispersion points to the relevance of Ni2.

\section{Summary and discussion}
Our calculations uncovered several important aspects of the La-327 correlated electronic structure in
the normal state. The Ni-$d_{z^2}$ dominated flat-band physics adopts different forms at ambient
and at high pressure. In the former case, with the help of additional strong renormalizations of
other dispersive states, it drives a SDW transition at $T_{\rm a1}$ which results in a 
repulsion of the flat-band character from the Fermi level in the emerging ordered state. This scenario
is qualitatively robust irrespective of the different layering motifs of 2222 or 1313 kind.
However, the 1313 motif adds layer-selective Mott physics to the problem which may further complexify
e.g. the detailed spin structure in the still to be determined ordered state. Thus low-$T$ ARPES so far
measured the ordered-state dispersions, and from our scenario one would predict that the flat-band
dispersion should shift back to the Fermi level with rising temperature towards $T_{\rm a1}$.
This mechanism with temperature has indeed been observed by ARPES for the trilayer nickelate La$_4$Ni$_3$O$_{10}$~\cite{lizhou17}, which displays a similar low-energy electronic structure.

Putting togther the previous results for the 2222 case~\cite{lechermann23} and the present ones for 1313, also the main feature of the high-pressure regime is apparently structure independent: the SDW transition should
be surpressed and the flat-band character right at the Fermi level remains intact until the superconducting transition takes place. This speaks for a key participation of the flat band in the formation of the superconducting state. Furthermore, the flat-band driven non-FL character is seemingly very strong in the 2222 high-pressure case, rendering Fermi-surface dispersion exceptionally weak~\cite{lechermann23}. Yet the additional layer-selective freedom in 1313 enables a concerted effort of the multiorbital electronic structure to counteract on the destructive low-energy flat-band fluctuations. This constructive-deconstructive dichotomy leads to the emergence of a manifest nodal-antinodal differentiation in the form of Fermi arcs. Such a strongly {\bf k}-selective spectral feature in the high-pressure regime could e.g. be detected by Raman spectroscopy~\cite{auvray21}.
If these Fermi arcs are a compelling companion
of high-$T_{\rm c}$ superconductivity in nickelates as they are in cuprates, or if they are a simple
bystander, has to be explored in future works.

\section{Acknowledgements} 
The work is supported by the German Research Foundation within the bilateral
NSFC-DFG Project ER 463/14-1. Computations were performed at the Ruhr-University
Bochum and the JUWELS Cluster of the J\"ulich Supercomputing Centre (JSC)
under project miqs.

\bibliography{literatur}
\end{document}